# Wood traceability system using blockchain and zero-knowledge proof

**K. Shibano**[1], T. Nakajima[2] and G. Mogi[1]

[1] Department of Technology Management for Innovation, The University of Tokyo, Tokyo, Japan
[2] Department of Forest Science, The University of Tokyo, Tokyo, Japan
E-mail: shibano@tmi.t.u-tokyo.ac.jp

**Summary:** The system proposed in this study uses zero-knowledge proof (ZKP) to verify the traceability of wood recorded in a public blockchain. Wood is a byproduct of several states, ranging from standing trees to logs, lumber, and wood products (hereinafter "wood objects"). The advantage of using the blockchain for record keeping is that participants can freely record the information at their discretion, without any restrictions. However, the openness of the blockchain may allow a malicious third party to introduce disinformation. In this study, we employ ZKP and near-field communication (NFC) chips to eliminate the possibility of disinformation introduction. ZKP is used to prove/validate changes in the state of wood objects, and the unique nonce associated with that state is encrypted and recorded on an NFC chip. The nonce is concealed and id of the wood object is defined as hash value of this nonce. We developed a prototype system based on an Android application and an Ethereum smart contract. We confirm that wood traceability and verification can be performed using the prototype system.

## 1. Introduction

In this study, we propose a traceability system for trees, logs, lumber, and final wooden products based on a public blockchain and zero-knowledge proof (ZKP). The blockchain has the advantage of allowing any user to keep records on it without restrictions. Furthermore, because records can be kept semi-permanently, it is possible to avoid the loss of existing tree records. However, because anyone can input a record on the blockchain, there is a possibility that third parties will record malicious disinformation. For example, if there is a very expensive tree, a person may wish to mislead others by claiming that his log was generated by that tree. ZPK can be used to verify which trees are used to make wood and which wood products are made from which wood, eliminating the possibility of disinformation. This can be done by using only the records on the blockchain. We have developed a prototype system using an Android application and an Ethereum smart contract to verify its operation.

End users will be able to confirm the origin of wood products using the proposed system. This may provide high added value that could not be realized until now. For example, a good-luck charm for academic success in school made from a tree on the campus of the same university may have a high added value. It would also help to reduce illegal timber.

## 2. Related studies

Several wood traceability systems that use blockchain have been proposed. Figorilli et al. (2018) use RFID, the blockchain, and a client-server application to implement a wood traceability system [1]. Cueva-Sánchez et al. (2020) propose a system that uses Hyperledger Fabric to eliminate illegal wood in the wood supply chain. They developed web and mobile applications [2]. There exists wood traceability system using not only blockchain but also ZKP. Xue et al. (2022) propose ZKP for a public blockchain-based system to prove transactions while ensuring privacy protection [3]. Baliyanet al. (2021) propose a highly transparent system that utilizes blockchain and RFID for general supply chain management systems. It prevents fraud by having the Law Enforcing Agency assess transactions. They mention wood traceability as an area of application [4]. Further details on blockchain-based wood traceability systems can be found in He and Turner (2022) [5]. The novelty of this study is that it uses ZKP to prove traceability. Traceability can be verified by the information in the blockchain only.

## 3. Zero-knowledge proof

ZKP is a protocol that allows a prover to tell a verifier that a proposition is true without conveying any knowledge other than that the proposition is true. We use zkSNARKs, a noninteractive zero-knowledge proof protocol used in many blockchain applications. The process to be proven has inputs and outputs and is converted into a circuit. Then, a trusted setup ceremony is performed to generate proving and verification keys. The prover generates a witness using the circuit, the proving key, and input. The verifier can confirm that the prover used the correct value for the private input using the verification key for the proof and public output. The public output is the output of the process and the value of the public input.

## 4. System overview

In this system, historical state records of wood supply chains, such as trees, logs, lumber, and wood products (hereinafter "wood objects") can be verified by referring to only blockchain records. A supply chain record has a tree structure and the state changes in one direction. We assume there are two users of the proposed system: a prover and a verifier. The prover is a wood object producer or processor, and the verifier is a consumer. The prover uses an Android application to record unique information of the wood object on a near-field communication (NFC) chip and generate a proof of ZKP. The NFC chip is attached to the corresponding wood object, and a proof of ZKP is simultaneously recorded to the blockchain when information is recorded on the NFC chip. The verifier can verify the wood object's traceability by verifying the proof. When writing to the blockchain, the signature is also recorded, allowing verification of who wrote the record. The key pair of the private and public keys of elliptic curve cryptography is stored in the Android application and can be used for signing and encryption/decryption. The Elliptic Curve Digital Signature Algorithm (ECDSA) is used for signatures, and the Elliptic Curve Integrated Encryption Scheme (ECIES) is used for public key cryptography.

## 5. Design of Android application and developing environment

The prototype system comprises an Android application, an Ethereum blockchain, and an NFC chip.

In this system, we use circom and snarkjs [6] for ZKP as libraries to implement. snarkjs [6] is used in the Android application to generate proof. The circuit and proving key data loaded in snarkjs are generated previously in the PC using circom, whose ZKP scheme is Plonk. Since it is a JavaScript library, it cannot be run directly in the application. A web server is set up within the application and accessed via WebView. web3j [7] connects to the Ethereum blockchain.

Key pairs associated with Ethereum's externally owned accounts are used for keys related to ECDSA and ECIES. The private key is stored in the application's storage area, bouncycastle [8] is used as the ECIES library, and web3j is used for the ECDSA library.

The Ethereum smart contract only records data for which the ECDSA signature and the proof of ZKP have been verified, and the ZKP verification contract is the snarkjs output.

The development environment is Ryzen 3600, 16 GB RAM (Windows 10), the Android device is Pixel3a (Android 12), the Ethereum blockchain is built locally using Ganache [9], and the NFC chip is MIFARE Classic 1k. Fig. 1 shows the Android device and NFC chip used in the development.

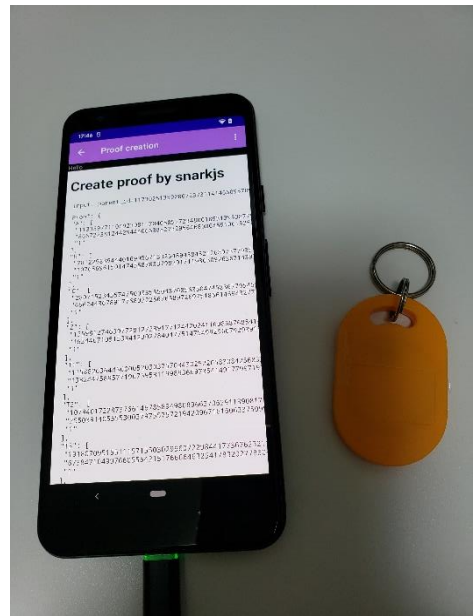

**Fig. 1 Android device and NFC chip**

## 6. Proof of traceability by zero-knowledge proof

A random number called "nonce" is encrypted and recorded on the NFC chip with its id. Each wood object $w$ has a unique id created, as expressed in (1),

$$id_w = hash(nonce_w). \qquad (1)$$

Ids are used to identify wood objects and are related to other metadata in or outside the blockchain. When running a proof/verification process, an error occurs if the id and nonce of the previous wood object state $p$ is not available. In this process, the public input is the id of $p$, the private input is $p$-nonce and $w$-nonce, the main process is the calculation of the hash value, and the output is the id of $w$. The flow of the process is shown in (2). If $w$ is a tree, $p$-nonce is assigned to 0, and $p$-id is assigned to a hash value of 0. Process (2) is converted into a circuit using circom [6]. Proving and verification keys are generated based on the circuit. The circuit data and the verification key are built into the Android application. The verification process using the verification key can be performed using an Ethereum smart contract.

```
function CalculateID (
  public input p_id,
  private input p_nonce,
  private input w_nonce) {
  p_hash = hash(p_nonce);
  p_eq = p_hash == p_id;
  w_hash = hash(w_nonce);

  return w_hash * p_eq;
}
```
(2)

After a nonce is generated using the prover's Android device, it is recorded on an NFC chip and then discarded. ECIES encryption is performed using the public key in the device, and the encrypted nonce is recorded on the NFC chip with the id. Therefore, once a nonce is written on the NFC chip, only the prover can decrypt it by reading the NFC chip. To generate an id of wood objects without trees, the parent's nonce is required. The device that recorded prestate in the NFC chip can read and decrypt nonce on that NFC chip. That nonce is received separately from that device, and along with its generated nonce, the public output and proof are generated in the process (2) and recorded in the blockchain. A QR code is used for transmitting the previous id and nonce between Android devices.

The sequence of all process is shown in Fig. 2. This shows an example that a log is generated from a tree. The first row shows initial setup procedure on PC. Output results are a circuit file and proving and verification keys. The circuit file and proving key are built-in Android application. The verification key is used for smart contract in the blockchain. The second and third rows show how to record the information of the wood objects in the blockchain using Android application.

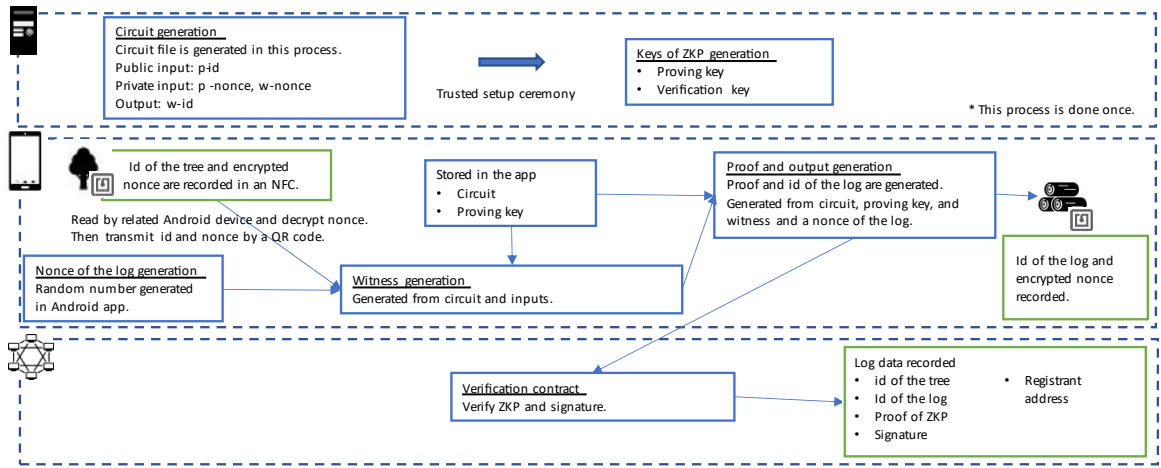

**Fig. 2** Schematic of the process flow, from tree to log

## 7. Prototype experiment

We confirmed that the developed prototype system works correctly. We have simulated a scenario where we cut down a tree and generated a log from it. We conducted two tests: (A) whether the NFC chip can be attached to an actual tree and remain without peeling off over a long period of time and (B) whether verification using the application worked properly.

(A) To check whether it is safe to leave the NFC chip on the tree, it was taped to a tree in a forest managed by the University of Tokyo. We confirmed that it remained there for 6 months without incident (Fig. 3).

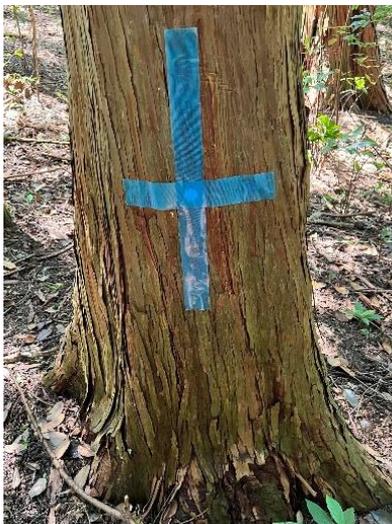

**Fig. 3** A tree with an NFC chip

(B) Next, we confirmed that only records verified by ECDSA signature and ZKP are recorded in the Ethereum blockchain. There are two Android devices: one for the tree and another for the log. The device for the tree read data from the NFC chip, decrypts nonce, and transmitted the tree nonce to the device for the log by a QR code on the app. The log device generated a new nonce and id of the log. Then, this device generates an id and an encrypted value of the new nonce to a new NFC chip. The proof and public output were recorded in the blockchain.

**Table 1** shows data recorded in the blockchain. Each record contains $w$-id, id of its previous wood object, or $p$-id, the proof of ZKP, ECDSA signature, and address of Ethereum. In the first row, the value of $p$-id is a hash value of 0. $w$-id of the first row and $p$-id of the second row is the same. For these two records, the verification of ZKP proof and the ECDSA signature was confirmed in the smart contract, proving that the nonce value of the previous wood object is known to the recorder when generating the second line's record. The ECDSA signature guarantees that the owner of the address performs this process. In other words, it verifies the traceability of a tree object and the recorder's address using only the information recorded in the blockchain.

| $w$-id | $p$-id | Proof of ZKP | Signature | Registrant |
|---|---|---|---|---|
| | | | | |
| 370043 348918 027287 066100 639847 781847 103961 878055 352655 234170 691329 159298 675 | 117302 513592 867237 311414 660957 099014 501703 690945 782888 424869 790425 860339 22425 | 0x1442dc394b656ea374badd091dfbfd97f4f33dd5169a2011024 fa6cb572dd1f417ac1a4c6cd7a9116a1db945ef7abe6da4f1f28d2 4fa950028032d6a3cdbee652e48db375c34b66021969ff6e59478 80cedc60dbc143eafb3ccbc56c0bb9bd7e18fdac8731746f68fb98 1bcaf336e932292c6ed4651cd3051595af1a466331e214e1e92e2 2d440fa39f862a70eab3c9d185ab2b418cf1240b0153d7a8cea5a 2a2b429e4f4566234b4d992b4ebc5cb004ed9bf90af689669e297 00fb2dcb0534915a2f7374c96da26ec3e6ab403e4eea885a07fed 8ba3e7add1936665fc534d91209 5dd0709c363b47fa00cd6939f 69883ed7e0dac90f77f9e95a019816a778 91da585e5f7d4fdc1dc 5897a85cb8c087c7f68ab4adebff9259696d4cb19513c11c780ed e61a29d7e0bb6caf55201a45e128cc47f885a8c36e7370833b4c7 18ba11e5ffea03e478971 52e1eac6487a9a5ffccf6243b872e18b9 aa6c86b835636f05204553de4fdbf0e1e478c26eebe4896351ab2 c1e29f572401a2aa35b4def9e2cb7fb242a833ff1bb5ae94f35b7b 02e9146fe05a08e3dedd2ba57765cff7114244b6e06df88b453b0 2fbf9766e7aae5845602 4c8f47ab00297998e5a335b8672ddde1d 14a9ffb17fcb63446a9126fca4b84da2b95f2a43cd1c40a2d1cd45 0c904a41568c4aa4779a0bdf92f744ea75602c9c5ba55a1c682d7 b1b762edec9e141799554a867d1226a33296333debc223a50e7d 2265ca2459a704b8222c8439611e5fe6265255313042 0a8ba66a6 c9b8be47b3afc05a201df8a961bbb99433b5fc18288608379a629 35f042e51bf3bd72c20443721bbb52ba3dfe177932aaa06da714ed a9e3e08a3f59bc6f28059b33b3239e628eb70aa2b4a2634ce3662 c6fe66e24c8fc8e3c4ad92b81d2dd659ac348b2726bd52aa1d73b fc1f2d323b6cc8145220acbf20046736fc75a6bf0bf984f2cb9df4 97b15547b534ae2cf8bf54c9ee6a0abc28a8463ef19d93b6a1f5eb 5c43892f167467623417a6a9c1f28f79687e011d63b48d3c3d3b 38829db2306d0ae9bdfd6a3213e67ac650293a8f04c66f4807040 34e0b0cf764e08cd557828063b00e2c98180d9eb829dc3c43db9 1172b9a59 | 0xa9f3dc7267 c499aa7b435b 92e0896374a6 c9cae2af279c8 4fdb590b7cfe5 ee894d9135ec e3ccf016efcb3 4c7ded4ccec9f 51dbb4d43cdc 8a78ec28f055e e1ff11c | 0xed883ff9c af287b84e4a de7d3d5587 d5532b7e4e |
| 190607 493880 360917 070689 790927 013619 637925 799095 905713 116413 289679 611404 98538 | 370043 348918 027287 066100 639847 781847 103961 878055 352655 234170 691329 159298 675 | 0x23fb2a027ead674e5a7dd9d48f667974d8764f9380e5cd1f4c6 34f5e8824e7ee110e754f75216b6d066b3faa3a10f31a3aba4613f 09302e5c0514752dc6927bb17c3c0f2be90dbcfabaefbae0cfc182 4a7f512bacd5bb35e6d41953172a91492026b57898e05de01f08 4b9fb179b40d33068e8fbb2ff61140991a6f4b1ae54a21ae04d2d 15d46cf70134125f588cb772df6c1535d133efc53c5690fd9bcbbf 41257e32482cb63b1c89b78885674ea96f8bf40dc96ac7039cbcc 16f5df1dfda3d1a0bfa19769da0b9ed5622 6ae7359bf74d937ebc 420669404 8ce38f182b998ec2c9918d649dbce7c82b72483b0bc 0e8cb695053422ca17 93183dac4ad3e500952f623 44ae0a8c322c e2cb8bd1450bb99680 60ff6a71b7b665d6a4501c4f09a1a01e570 1dfd5972e8 badaa7c256f40cd9bfacd5201d1b0de8ab6d6cea2c9 17eee12f77396daeb3c084f66899c348989e9f482178e78d0f47d 8fc4b21475608fe31aa0fe53570ea50f5ebd666777f5512fdcf329 15d8c5b7fd4f836746b06524ca02abe57eb712991ae187b2e742f fba1c12d3c66350850b08cccce2bf49cb5a22fb5f7fefd2c8bba7 9e566687ee524b801d64e3ff3ca44baffd6f58e2dde58a2268e04c 8774f15fdf4948c4d5ef6cebd004d51361ceaa31ef1446ec7ee021 f731cdc4a92e229f59562072201c956c38f23cd58e90876474b8d e4b1ece0279ec614334ac094d00151c080b41581b69cd48fa145 ea555501bf00e48bc7232e0e11282241d558c755bad13a3ad0ea 6fc6efb42cde45f2952a1e38ce992c8f20abd910055800297dffb9 2499f8ccd5d5044a75bfe01a9212f24536d342264d3fbb85277bb 96a424 1d4e22d8fa9 79b917a b8e115ca5be3877e2bc81cf10509 9dea8d07897114067307777f086cd6541e6d78407e4429051f42d 16ef899c83f2066cb62310f24124efcc0ca7e8c9f9ec93e411behd 9b9458c4cbd31c5d6a469f5f70131728990a868e5c619b1ebc32c 219eb067b150b4dca8b8b62f7f76fbadc1a009627921d895346c2 9add236c77175bfca19fc248055b0fb1a2f5949af49677b7b2260 2a88d26e89d229407309526a9b9d803e9515db965f917f32a131 f1421a3e1 | 0x68ce9d0326 660610bbccdc 3767feac7f7a2 3203c2df8898 1f4bfc5f9a07f 421c35d908d9 c3f771984c90 e7563f6bab f43 1b193bf121a0 db08a1f3927f9 8967e41c | 0xab00c6d6 2d94976e50 6d602c3190 8fb89d8651 e2 |

**Table 1 Examples of records recorded in the blockchain**

Generally speaking, ZKP proof generation requires much computation and is time-consuming. In the circuit of this study, the number of constraints is 731. We measured the time required to generate proof on an Android device and found that the average time for ten trials was 5,858.6 milliseconds. Pixel3a is a middle-class device several years old, and this time is considered sufficient for practical use.

The gas consumption for the smart contract was 1,030,996 [gas], 10 trials on average, approximately 40.2 USD in terms of gas prices in October 2022 (1 gas = 30 gwei, 1 Ether = 1,300 USD) [10][11]. The transaction cost is high because of the many processes involved in the smart contract, such as verifying ZKP and signatures and registering information.

## 8. Future work
This study's system only records the traceability relationship of wood objects in the blockchain. However, it is crucial to record additional information associated with each $w$-id; for example, GPS location information, pictures of wood objects, and tree species. Adding such information to traceability will increase its utility. To make it persistent, we plan to record additional information in a distributed database, such as IPFS.

To reduce transaction costs, we consider using another blockchain compatible with EVM.

We should also confirm whether this system can be used without problems for actual lumber processors. We plan to assess this with Japanese companies.

Although this system is applied to the traceability of wood in this study, it can be applied to the traceability of all products with the same relationships. We study what products and goods the system can be applied to and what value it might generate.

## 9. Conclusions
In this study, we proposed a method for verifying wood traceability using blockchain, NFC chips, and zero-knowledge proof. We constructed a prototype system and confirmed that wood traceability verification can be performed accurately. This system is a sample application of ZKP using an Android application and the Ethereum blockchain, and we hope this study's results will help develop applications using ZKP.

**Acknowledgments**
This work has been supported by Endowed Chair for Blockchain Innovation and the Mohammed bin Salman Center for Future Science and Technology for Saudi–Japan Vision 2030 (MbSC2030) at The University of Tokyo.